\documentclass[11pt, a4paper, Physsubmission, Phys]{SciPost} 

\hypersetup{
    colorlinks,
    linkcolor={red!50!black},
    citecolor={blue!50!black},
    urlcolor={blue!80!black}
}

\usepackage[bitstream-charter]{mathdesign}
\usepackage[export]{adjustbox}
\DeclareSymbolFont{usualmathcal}{OMS}{cmsy}{m}{n}
\DeclareSymbolFontAlphabet{\mathcal}{usualmathcal}

\begin{document}

\begin{center}{\Large \textbf{
India Based Neutrino Observatory, \\ Physics Reach and Status Report \\
}}\end{center}

\begin{center}
D. Indumathi\textsuperscript{1},
\end{center}

\begin{center}
{\bf 1} The Institute of Mathematical Sciences, Chennai and Homi Bhabha
National Institute, Mumbai
\\
indu@imsc.res.in
\end{center}

\begin{center}
\today
\end{center}


\definecolor{palegray}{gray}{0.95}
\begin{center}
\colorbox{palegray}{
  \begin{tabular}{rr}
  \begin{minipage}{0.1\textwidth}
    \includegraphics[width=30mm]{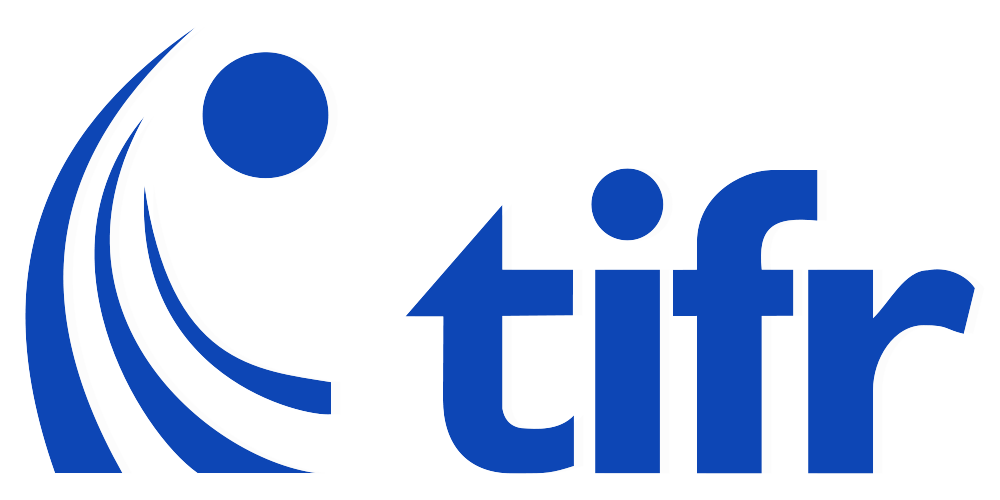}
  \end{minipage}
  &
  \begin{minipage}{0.85\textwidth}
    \begin{center}
    {\it 21st International Symposium on Very High Energy Cosmic Ray Interactions (ISVHE- CRI 2022)}\\
    {\it Online, 23-28 May 2022} \\
    \doi{10.21468/SciPostPhysProc.?}\\
    \end{center}
  \end{minipage}
\end{tabular}
}
\end{center}

\section*{Abstract}
{\bf
The India-based Neutrino Observatory (INO) is a proposed underground facility located in India that will primarily house the magnetised Iron CALorimeter (ICAL) detector to study atmospheric neutrinos produced by interactions of cosmic rays with Earth's atmosphere. The physics goal is to to make precision measurements of the neutrino mixing and oscillation parameters through such a study. We present here the results from detailed simulations studies, as well as a status report on the project. In particular, we highlight the sensitivity of ICAL to the open issue of the neutrino mass ordering, which can be determined {\it independent of the CP phase} at ICAL.
}

\vspace{10pt}
\noindent\rule{\textwidth}{1pt}
\tableofcontents\thispagestyle{fancy}
\noindent\rule{\textwidth}{1pt}
\vspace{10pt}

\section{Introduction}
\label{sec:intro}
The possibility of neutrino oscillations was originally suggested by
Pontecorvo (actually, he examined the possibility of neutrino
anti-neutrino oscillations) \cite{Pontecorvo:1957cp}.
The theory of neutrino mixing in
three-flavours was given by Maki, Nakagawa and Sakata
\cite{Maki:1962mu} and the
effect of matter on the propagation of neutrinos was discussed by
Wolfenstein \cite{Wolfenstein:1977ue},
Mikheev and Smirnov \cite{Mikheev:1986wj}.
It is now well-established that neutrino flavours mix, and hence
neutrino oscillations have been observed in diverse instances such as
in solar neutrinos, atmospheric neutrinos, reactor neutrinos, etc.; see
Ref.~\cite{Bilenky:2016pep} for a historical review. Neutrino mixing in
three flavours is parametrised by the PMNS matrix, $U_{PMNS}$, given by
\begin{eqnarray}
U_{PMNS} = \left(
          \begin{array}{ccc}
          c_{12}c_{13} & s_{12}c_{13} & s_{13}e^{-i\delta_{CP}}  \\
 -c_{23}s_{12} - s_{23}s_{13}c_{12}e^{i\delta_{CP}} & c_{23}c_{12} -
s_{23}s_{13}s_{12}e^{i\delta_{CP}}&  s_{23}c_{13}\\
  s_{23}s_{12} - c_{23}s_{13}c_{12}e^{i\delta_{CP}}& -s_{23}c_{12} -
c_{23}s_{13}s_{12}e^{i\delta_{CP}} & c_{23}c_{13} \end{array} \right).
\label{eq:pmns}
\end{eqnarray}
Here $c_{12}=\cos\theta_{12}$, $s_{12}=\sin\theta_{12}$ etc., and
$\delta_{CP}$
denotes the CP violating (Dirac) phase. By definition, the 3 $\times$
3 neutrino mass matrix $M_\nu$ is diagonalised in the charged-lepton
mass basis.
The parameters involved are the mixing angles $\theta_{ij}$ and the
mass-squared differences $\delta_{ij} \equiv m_i^2 - m_j^2$. A combined
analysis \cite{Esteban:2020cvm}
of the data from various experiments indicates
that the across-generation mixing angle $\theta_{13} \sim 8.5^\circ$ is
smaller than $\theta_{12} \sim 34^\circ$, $\theta_{23} \sim 45^\circ$,
while the CP phase is still not well-established. There are two
independent mass squared differences, $\vert \delta_{31} \vert \gg
\delta_{21}$. While $\delta_{21} \sim 7.6\times 10^{-5}$ eV$^2$ is known
to be positive, the {\em sign} of the other mass squared difference is
unknown, with $\vert \delta_{31} \vert \sim 2.5 \times 10^{-3}$ eV$^2$,
with the sum of all the neutrino masses constrained by cosmological
considerations to be less than about 2 eV.
One of the important open issues to be settled by future experiments is
the so-called neutrino mass ordering, {\em viz.}, whether the third mass
eigenstate is heavier or lighter than the others.
We will discuss the proposed INO in this context.

\section{The INO project}
\label{sec:INO_overview}

The INO is a proposed mega-science project that is jointly funded by the
Department of Atomic Energy and the Department of Science and
Technology, Government of India. The immediate goal is the creation of
an underground (at least 1 km deep) cavern for scientific research
purposes, with the main ICAL detector to study atmospheric neutrinos and
cosmic muons. It will incorporate a centre for particle physics and
detector technology at Madurai, South India, which currently houses the
mini-ICAL prototype. The INO graduate programme trains students on both
the theoretical and experimental aspects of neutrino/particle physics.

\subsection{The ICAL detector}
\label{sec:ICAL}

\begin{figure}[htp]
\centering
\includegraphics[width=0.4\textwidth]{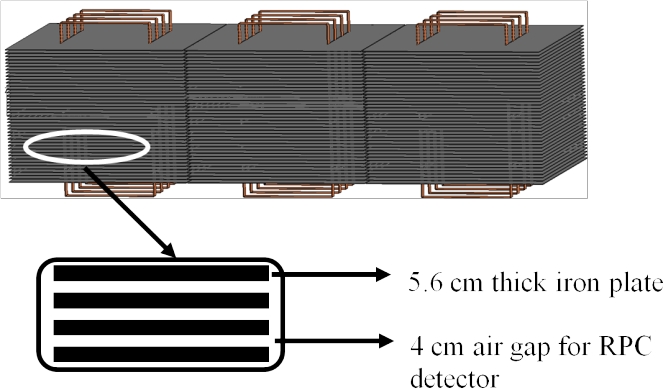} \hspace{1cm}
\includegraphics[width=0.4\textwidth]{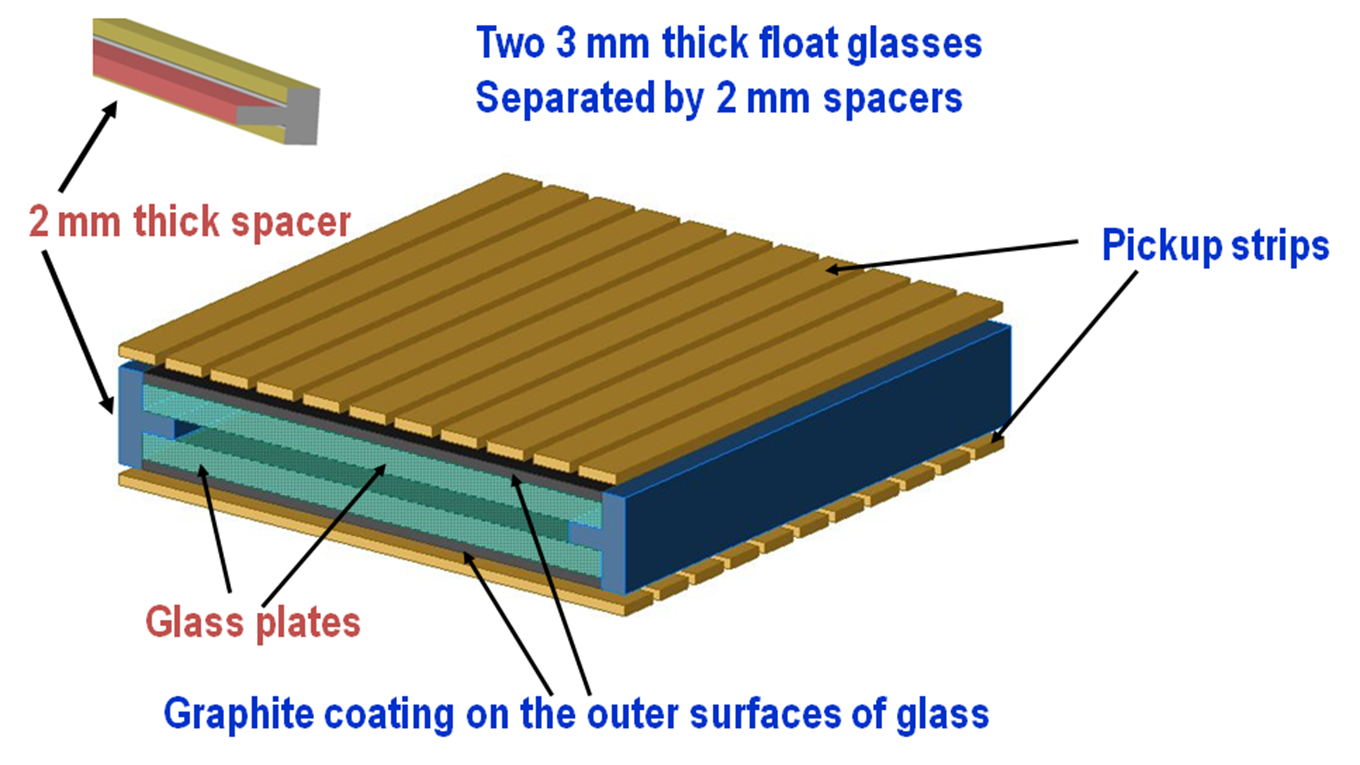}
\caption{Schematic of ICAL detector and RPCs.}
\label{fig:ICAL_schematic}
\end{figure}

The proposed \cite{ICAL:2015stm}
51 kton ICAL detector will consist of 151 layers of 56 mm thick
soft iron that can be magnetised to about 15 kGauss. It will consist of
three identical modules with the layers of iron plates separated by 40
cm air gaps in which the roughly 30,000 active Resistive Plate Chambers
(RPCs) of dimensions $2 \times 2$ m$^2$ will be housed; see
Fig.~\ref{fig:ICAL_schematic}. The high DC voltage across the 3 mm glass
plates separated by 2 mm of gas gap creates a discharge when a charged
particle passes through them. These are digitised and stored as hits
using the nearly 4 million channels of readout electronics, which are
later analysed to reconstruct the kinematics of each event. While the
minimum ionising muons leave clean long tracks in the detector, the
hadrons shower. Hence the momentum, including direction, of the muons
is well-reconstructed compared to that of the hadrons; in addition, the
sign of the muon's charge can also be reconstructed to better than 98\%
for the few GeV events of interest.

\section{Atmospheric Neutrinos and ICAL}
\label{sec:atm}
Cosmic rays interact with the Earth's atmosphere to produce muons, which
further decay to neutrinos, producing muon and electron neutrinos (and
anti-neutrinos) in roughly 2:1 ratio. These neutrinos will mainly interact
with the iron in the detector to produce charged particles in
charged-current (CC) interactions:
\begin{eqnarray} \nonumber
\nu_\mu + N & \to & \mu^- + X~, \nonumber \\
\overline{\nu}_\mu + N & \to & \mu^+ + X~,
\label{eq:CC}
\end{eqnarray}
where $X$ is any hadronic debris. The magnetic field bends the
differently charged muons in different directions and hence
differentiates between neutrino and anti-neutrino induced events. This
separation allows for a sensitive study of matter effects which are
different for neutrinos and anti-neutrinos and hence can pin down the
neutrino mass ordering.

\subsection{ICAL Detector Response}
\label{sec:det}

The physics reach of ICAL using detailed GEANT4 based simulations
\cite{GEANT4:2002zbu} is shown in this and the following sections. Unless
specifically mentioned, all results are from the INO white paper
\cite{ICAL:2015stm}; we give the main highlights here.

\begin{figure}[htp]
\centering
\includegraphics[width=0.6\textwidth]{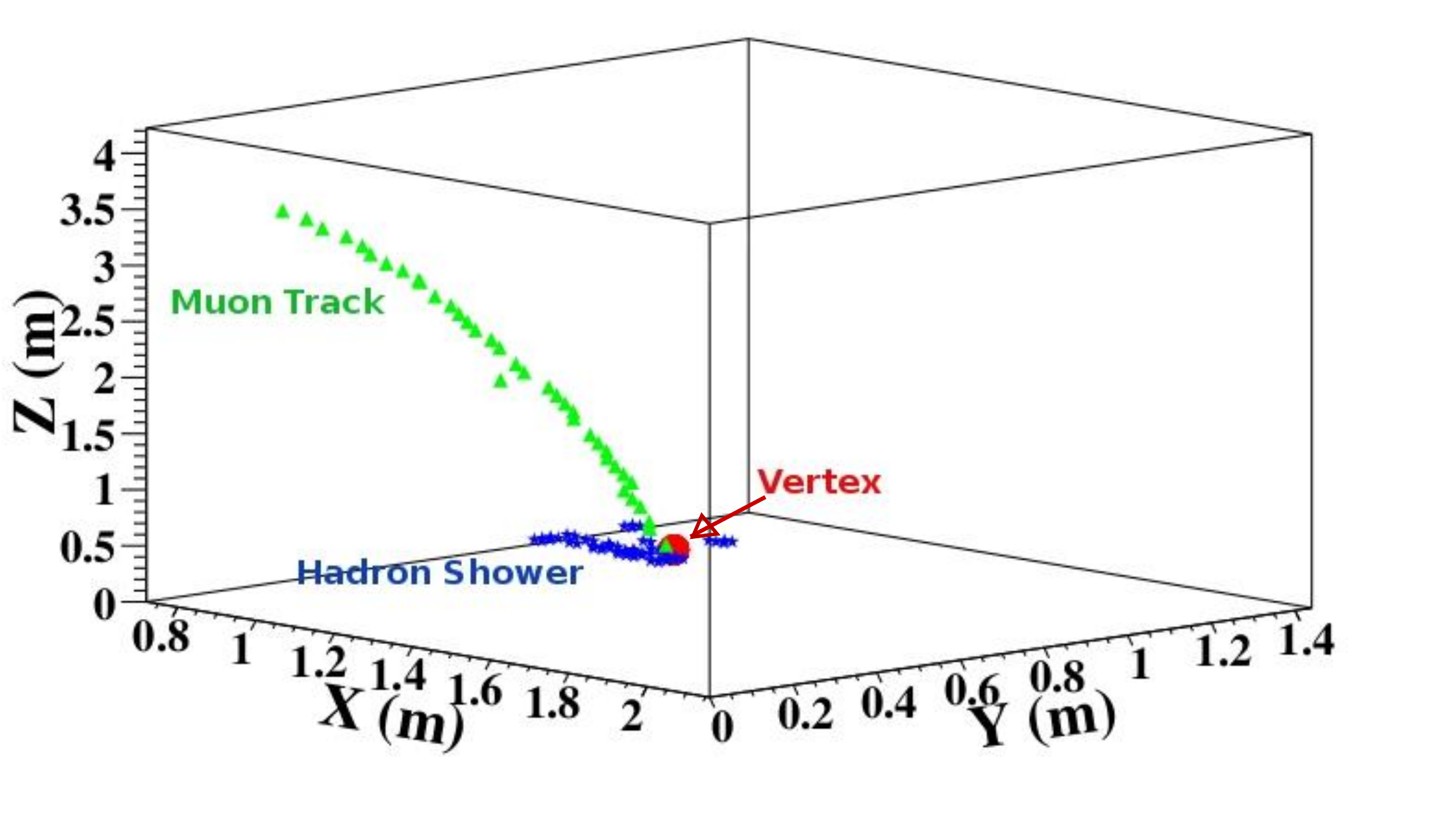}
\caption{Typical event in ICAL simulations.}
\label{fig:vice}
\end{figure}

The magnetic field in ICAL allows for a separation of $\mu^-$ and
$\mu^+$ (neutrino or anti-neutrino induced) events as well as helps to
determine the magnitude and direction of the muon's momentum. A typical
atmospheric neutrino event is shown in Fig.~\ref{fig:vice}. The lomg
tracks of the mimimum ionising muons curve in the presence of the
magnetic field and their momentum is reconstructed using a Kalman filter
algorithm. The hadrons shower and appear as hits clustered close to the
vertex. Typically 10,000 events with fixed momentum and direction were
propagated in the simulated ICAL detector with GEANT, and their
behaviour analysed.

\begin{figure}[htp]
\centering
\includegraphics[width=0.49\textwidth]{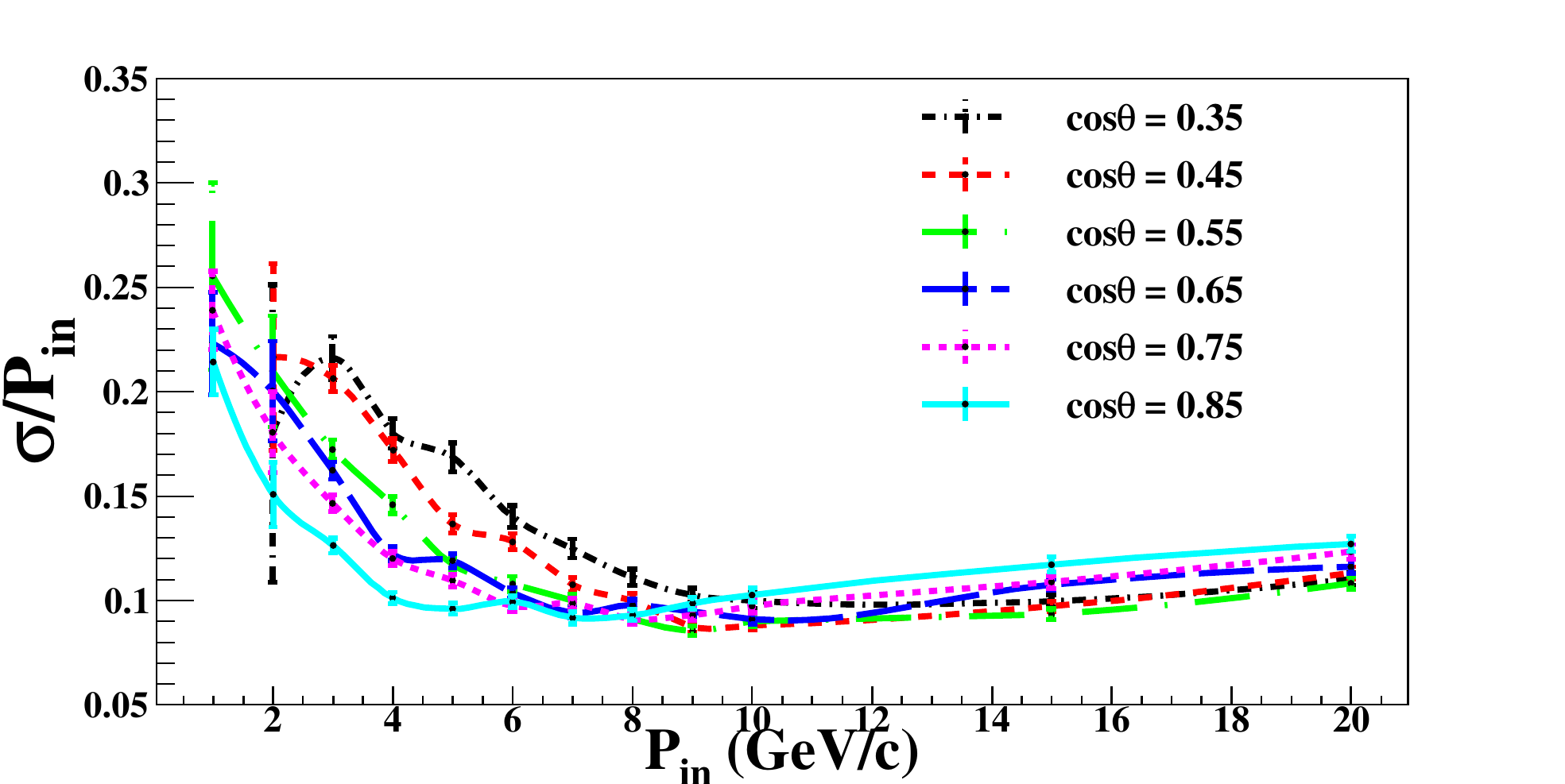}
\includegraphics[width=0.49\textwidth]{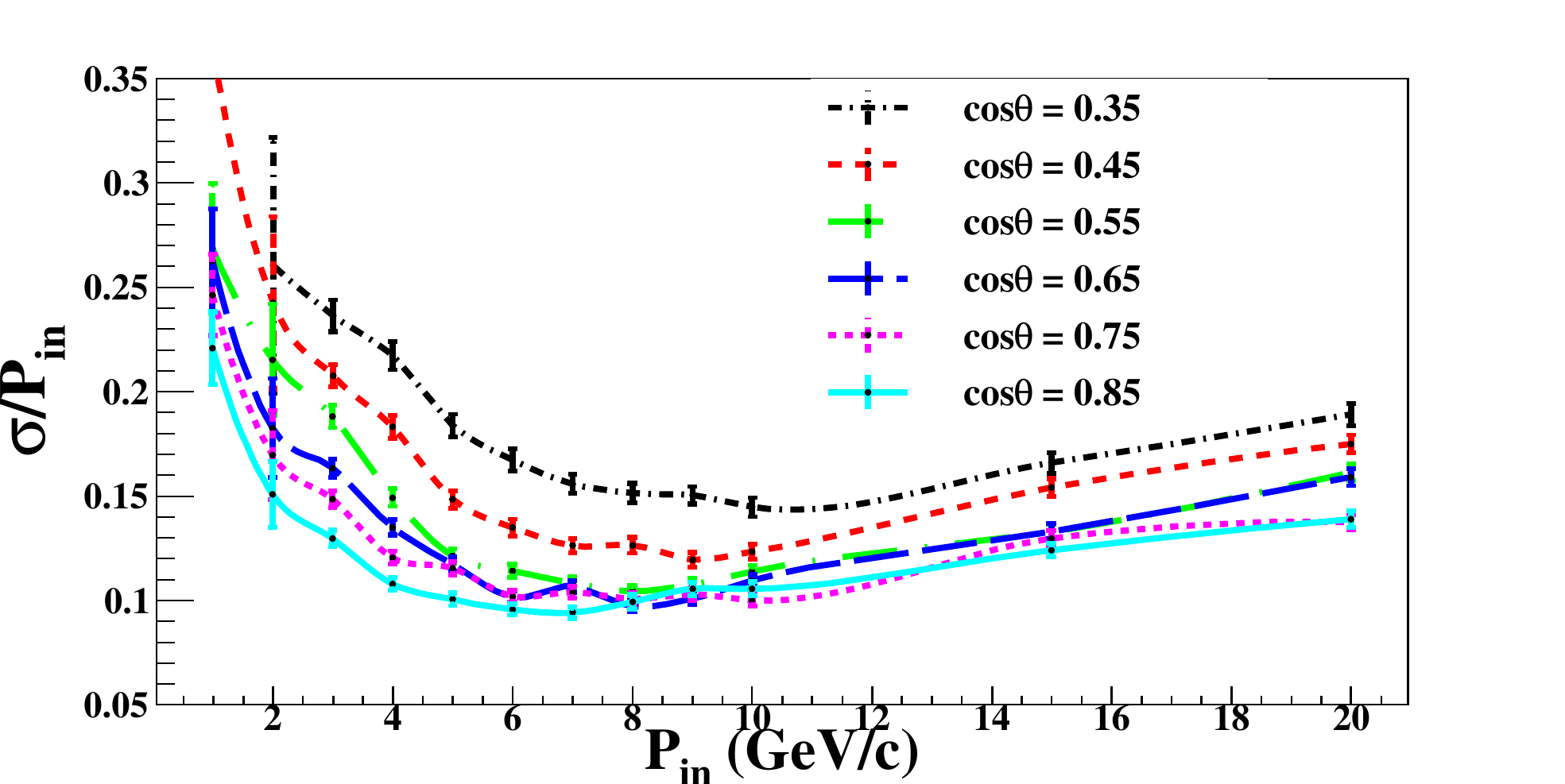}
\includegraphics[width=0.49\textwidth]{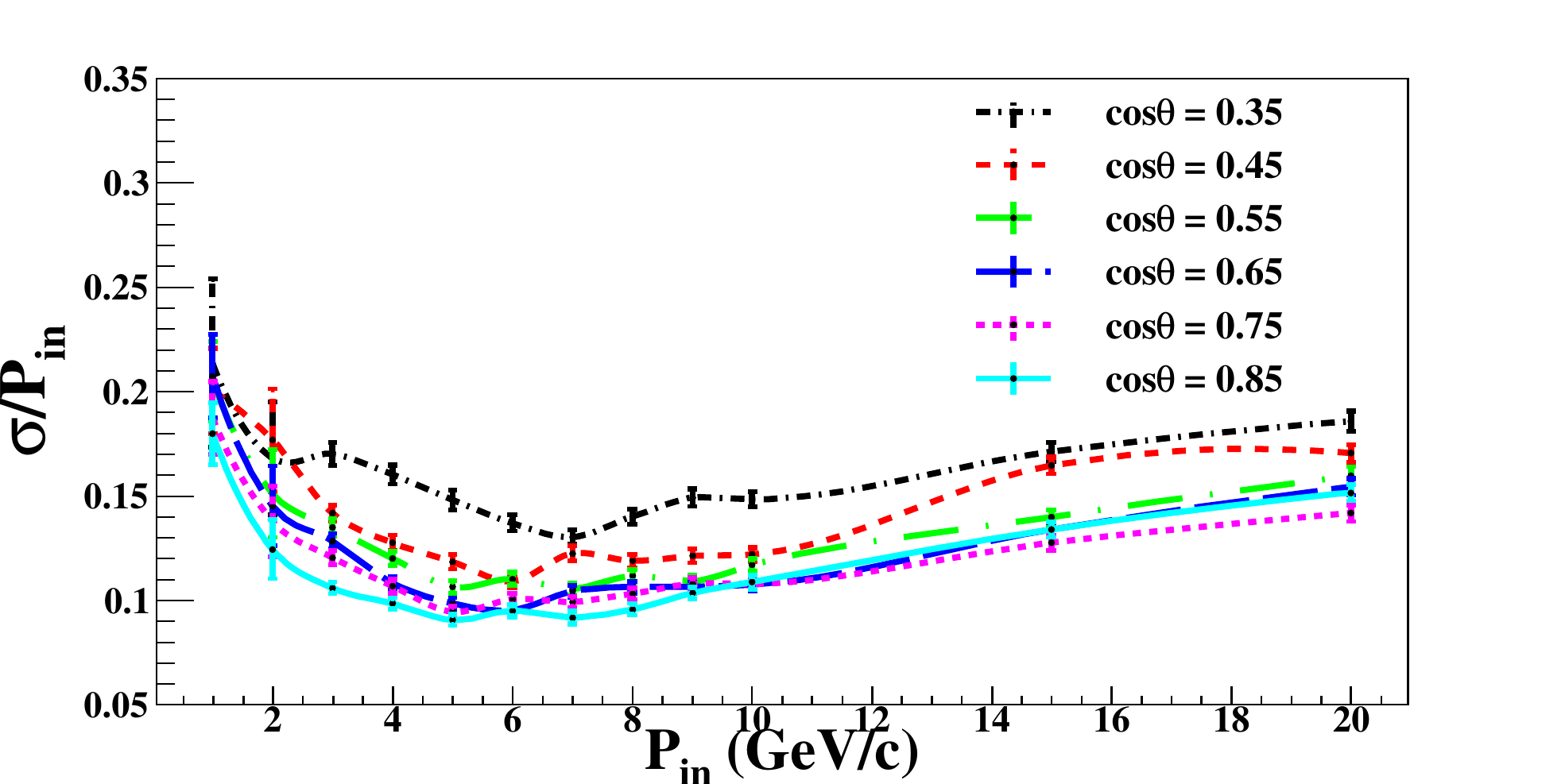}
\includegraphics[width=0.49\textwidth]{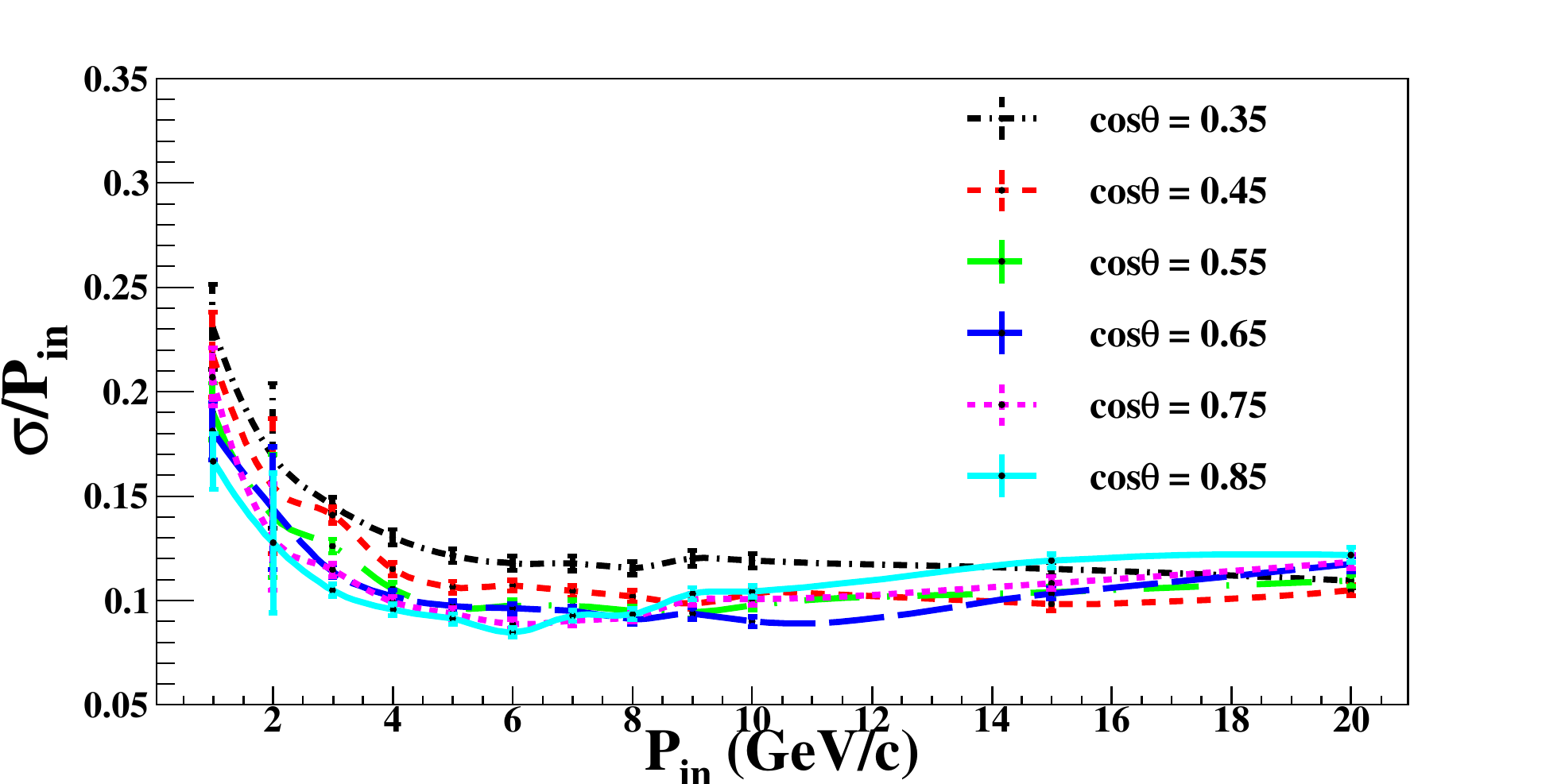}
\caption{Muon resolution as a function of input momentum and
$\cos\theta$, for different regions of azimuthal angles,
$\vert\phi\vert \le \pi/4$ (top left), 
$\pi/4 \le \vert\phi\vert \le \pi/2$ (top right), 
$\pi/2 \le \vert\phi\vert \le 3\pi/4$ (bottom left), 
$3\pi/4 \le \vert\phi\vert \le \pi$ (bottom right).}
\label{fig:momres}
\end{figure}

The muon momentum resolution (defined as $\sigma/P_{in}$, obtained by
reconstructing fixed momentum muons in ICAL and fitting the resultant
distribution to a gaussian distribution) is shown in Fig.~\ref{fig:momres}
for different values of $(\cos\theta, \phi)$, with the upward direction
corresponding to $\cos\theta=1$. The azimuthal dependence occurs
due to the presence of the magnetic field which breaks the azimuthal
symmetry. (In addition, the angular resolution (for the zenith angle)
is better than $1^\circ$ for muons with more than $\approx 2$ GeV energy.)

\begin{figure}[htp]
\centering
\includegraphics[angle=90,width=0.49\textwidth,height=0.3\textwidth]{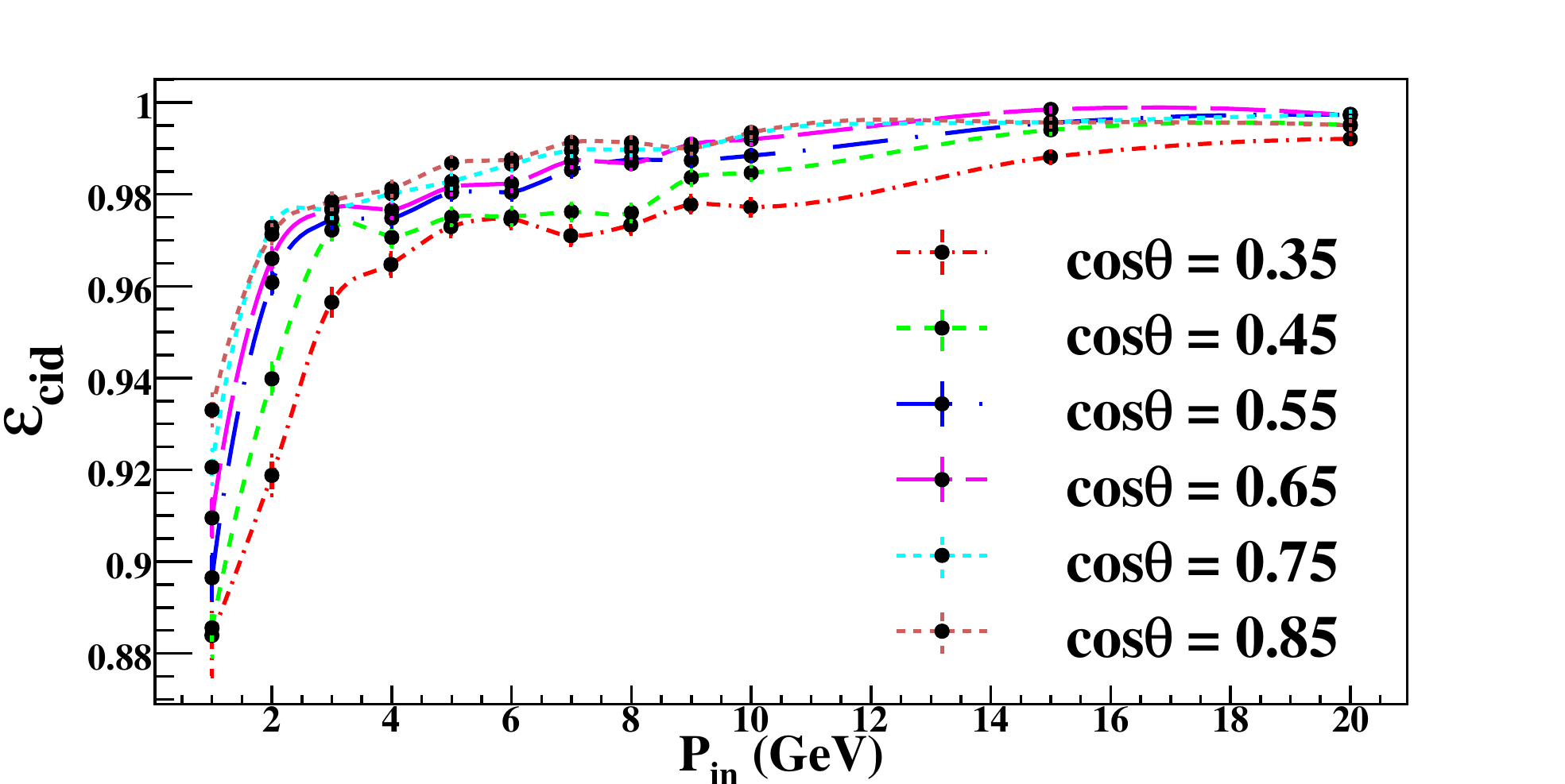}
\includegraphics[angle=90,width=0.49\textwidth,height=0.3\textwidth]{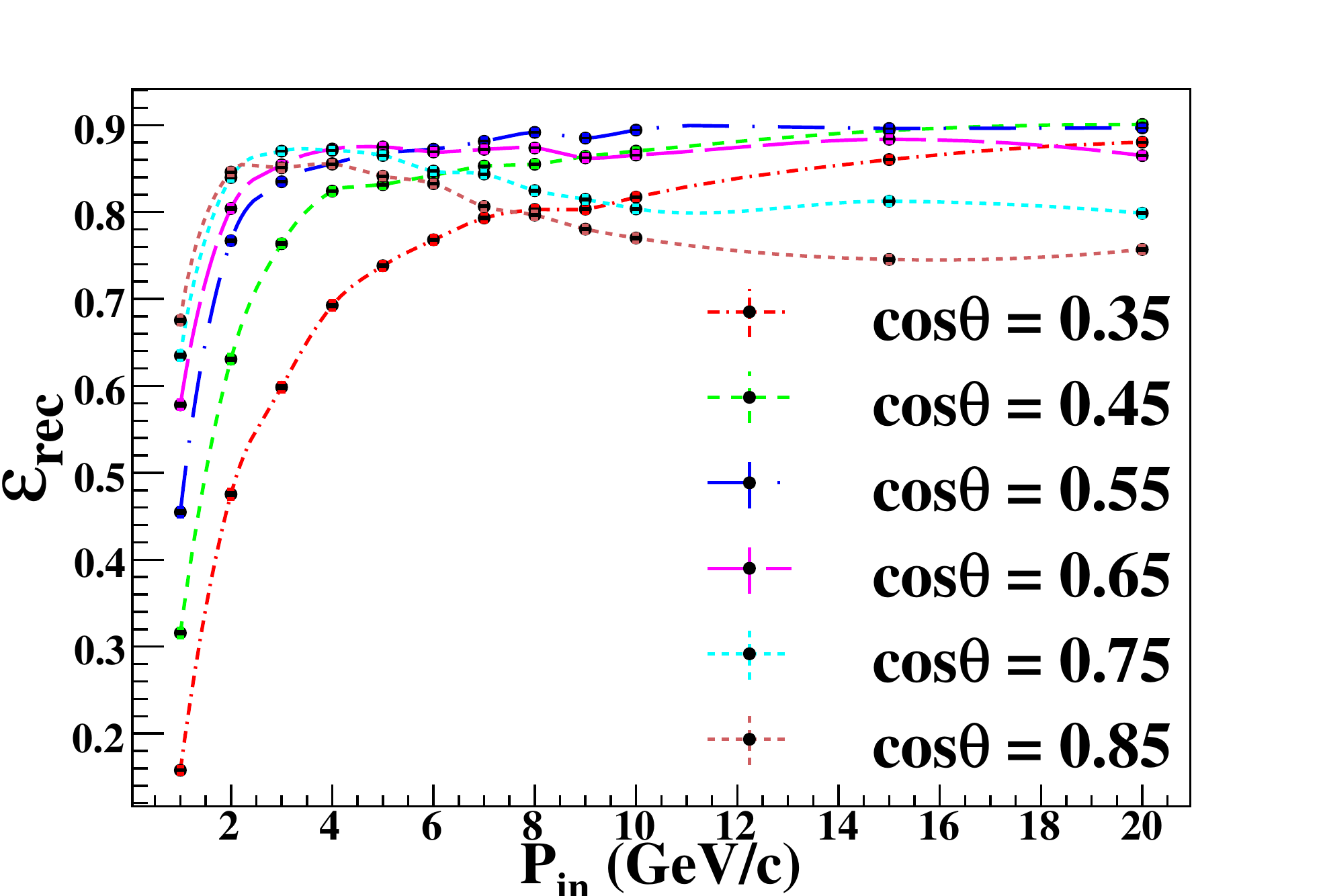}
\caption{Reconstruction efficiency $\epsilon_{rec}$ (left) and 
charge-id efficiency $\epsilon_{cid}$ (right) as a function of the input
momentum for different values of the zenith angle.}
\label{fig:eff}
\end{figure}

Fig.~\ref{fig:eff} shows the reconstruction $\epsilon_{rec}$ and relative
charge-id efficiency $\epsilon_{cid}$. The relative charge-id efficiency
is the fraction of reconstructed events that were reconstructed with
the correct charge sign. We see that $\epsilon_{cid} \gtrsim 98$\%
over a large range of momentum magnitude and direction.

\begin{figure}[htp]
\centering
\includegraphics[width=0.6\textwidth]{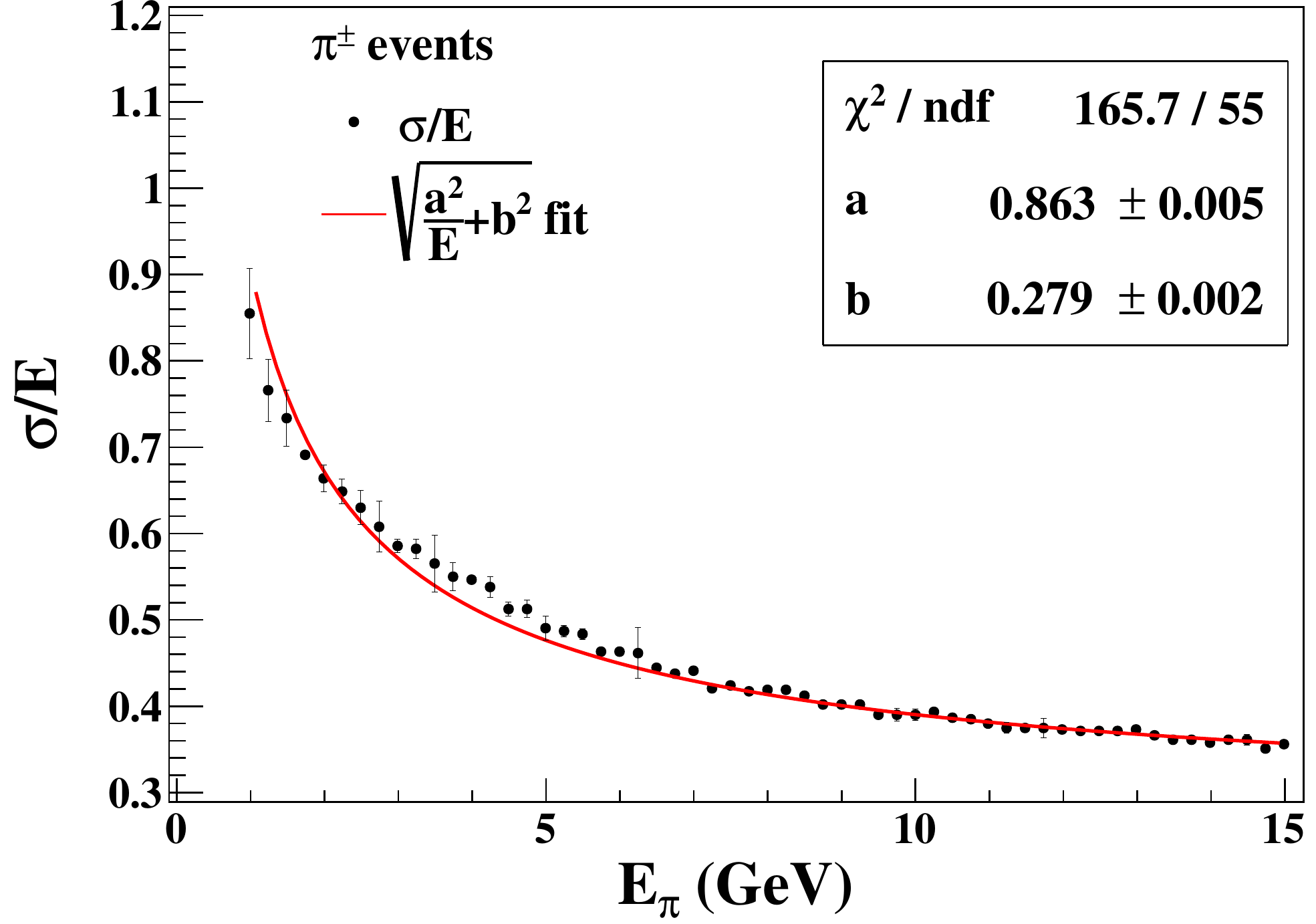}
\caption{Hadron energy resolution as a function of pion energy. Here the
reconstructed hadron energy was fitted to a Vavilov distribution and its
resolution $\sigma/E$ computed. See \protect Ref.~\cite{ICAL:2015stm}
for details.}
\label{fig:hadron}
\end{figure}

Finally, the hadron energy resolution for fixed energy charged pion
events is shown in Fig.~\ref{fig:hadron}. The final hadronic component
in neutrino-iron interactions of interest at ICAL can contain multiple
hadrons; however, pions dominate the sample; for details, please see
Ref.~\cite{ICAL:2015stm}.

\subsection{Precision measurement of the 2--3 parameters}
\label{sec:prec23}

\begin{figure}[htp]
\centering
\includegraphics[width=0.6\textwidth]{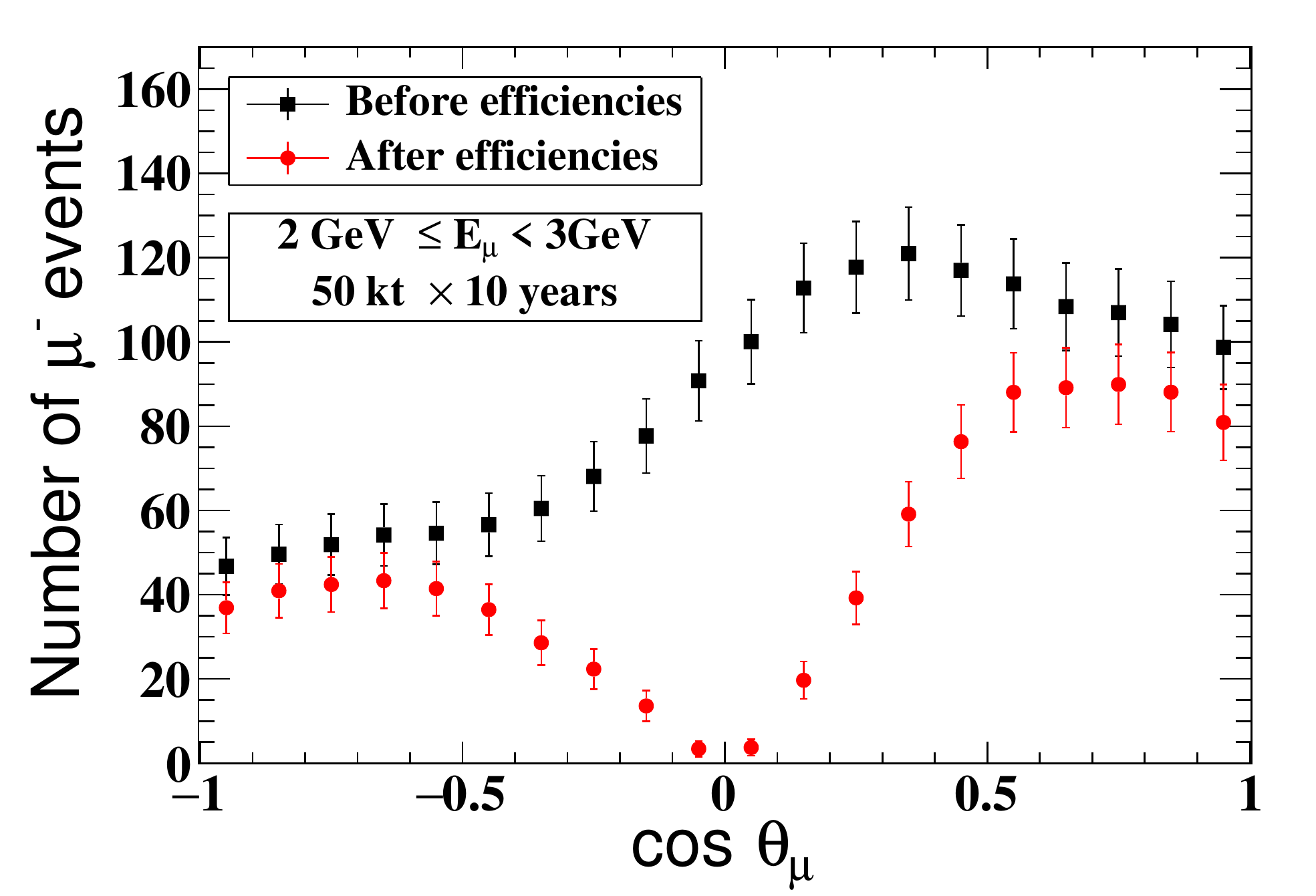}
\caption{Zenith angle distributions of oscillated atmospheric muon
$\mu^-$ events with energies from 2 to 3 GeV for an exposure time of 10
years. Statistical errors are shown. Note that there are about 2.5 times
fewer $\mu^+$ events due to the smaller cross sections although their
fluxes are approximately the same.}
\label{fig:events}
\end{figure}

Honda atmospheric neutrino fluxes \cite{Honda:2011nf} were used to
generate 1000 years of events using the ICAL geometry and the NUANCE
neutrino generator \cite{Casper:2002sd} and then oscillated and scaled
appropriately for the analysis. Typical events as a function of the
muon zenith angle $\cos\theta$ are shown in Fig.~\ref{fig:events} for
events in the muon energy bin 2--3 GeV with an exposure time of 10
years. Finite detector resolutions smear these numbers somewhat. Notice
the poor reconstruction in the horizontal direction.

\begin{figure}[htp]
\centering
\includegraphics[width=0.49\textwidth,height=0.3\textwidth]{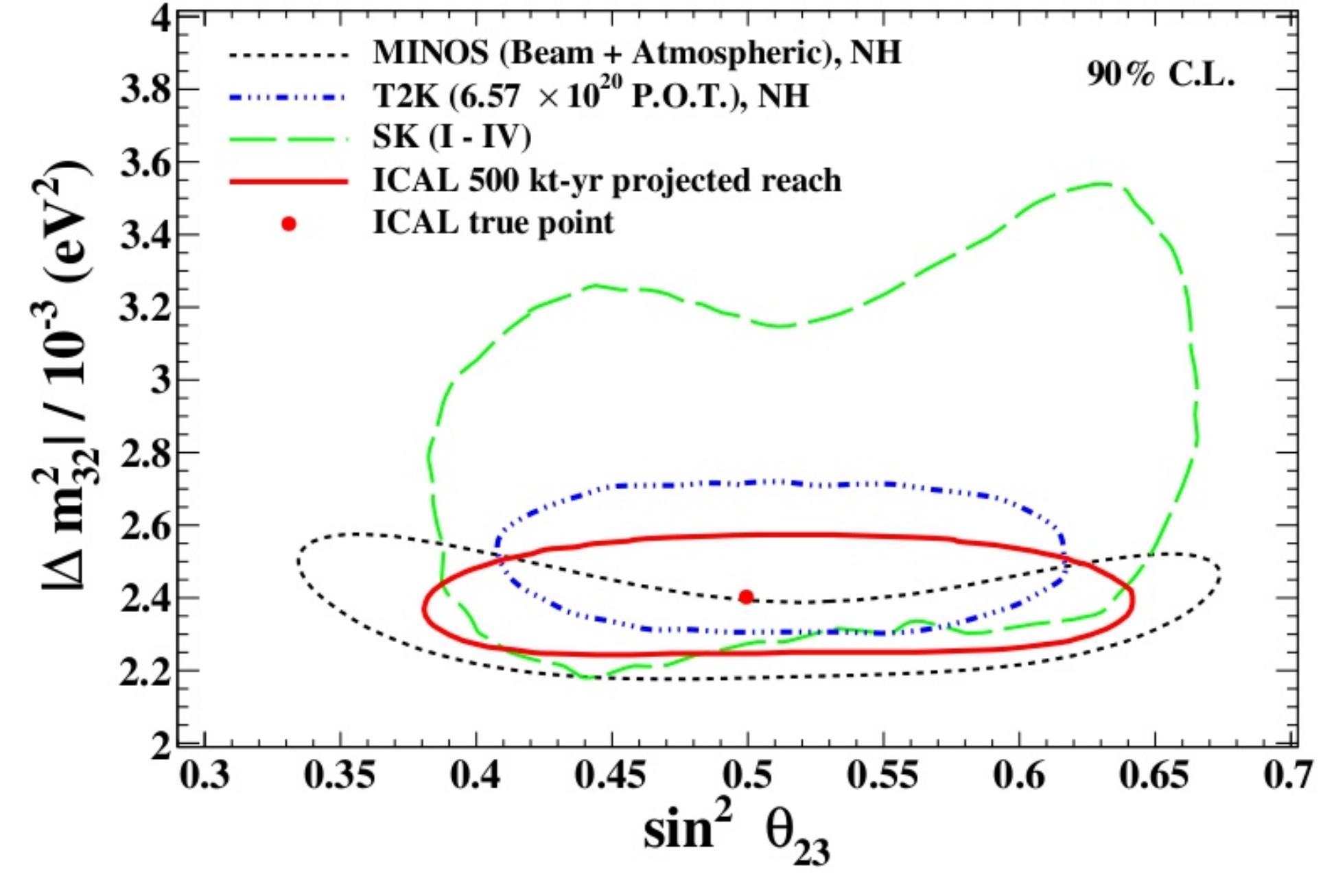}
\includegraphics[width=0.49\textwidth,height=0.3\textwidth]{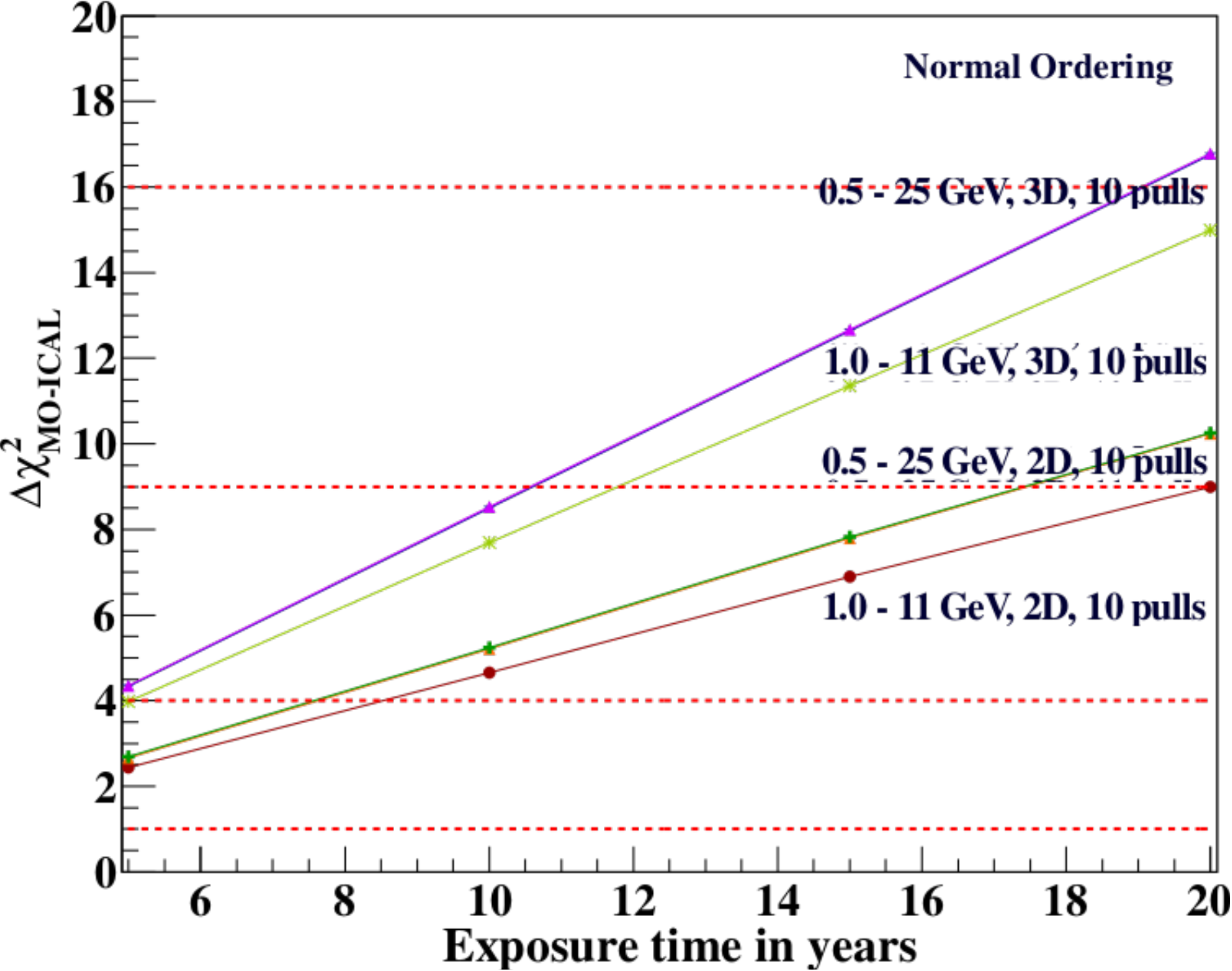}
\caption{Left: Precision reach in 10 years of ICAL for $\sin^2\theta_{23}$
and $\delta_{32}$, including hadron energy information. Right: Sensitivity
to the mass ordering for assumed input normal ordering as a function of
exposure time; similar results are obtained for inverted ordering. The
sensitivity increases with inclusion of larger muon energy range and
inclusion of hadron energy information.}
\label{fig:prec}
\end{figure}

Fig.~\ref{fig:prec} shows the precision that can be obtained in 10
years in the 2--3 parameters, $\sin^2\theta_{23}$ and $\delta_{32}$
for input values of $(0.5, 2.4\times 10^{-3}$ eV$^2)$ in comparison with
other experiments. Note that ICAL is yet to be built!  The results are
marginalised over the allowed $3\sigma$ range of $\sin^2\theta_{13}$
and include the reconstructed muon momentum and direction information,
as well as hadron energy information. Results without hadron energy
information are labelled {\sf 2D} and those with hadron energy included
are labelled {\sf 3D} in the figure.

Fig.~\ref{fig:prec} also shows the sensitivity to the mass ordering:
plotted is the minimum $\chi^2$ obtained when the data is fitted to the
wrong (inverted) mass ordering as a function of the exposure time when the
true ordering is normal. The result is marginalised over the magnitude of
$\delta_{32}$ as well as $\theta_{23}$ and $\theta_{13}$. The inclusion
of hadron information improves the result considerably. Similar results
are obtained when the true hierarchy is inverted.

\begin{figure}[htp]
\centering
\includegraphics[width=0.49\textwidth,valign=t]{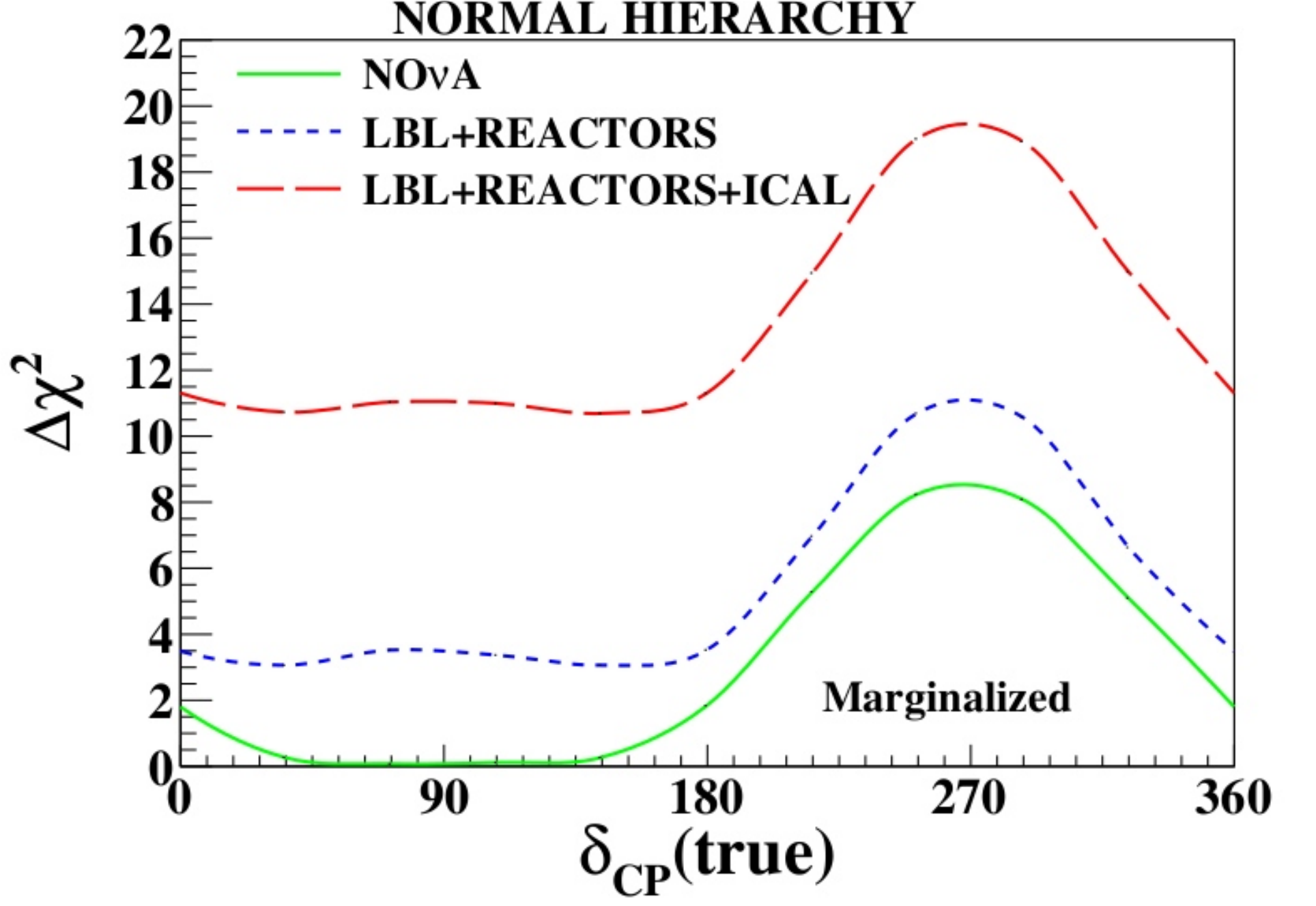}
\includegraphics[width=0.49\textwidth,valign=t]{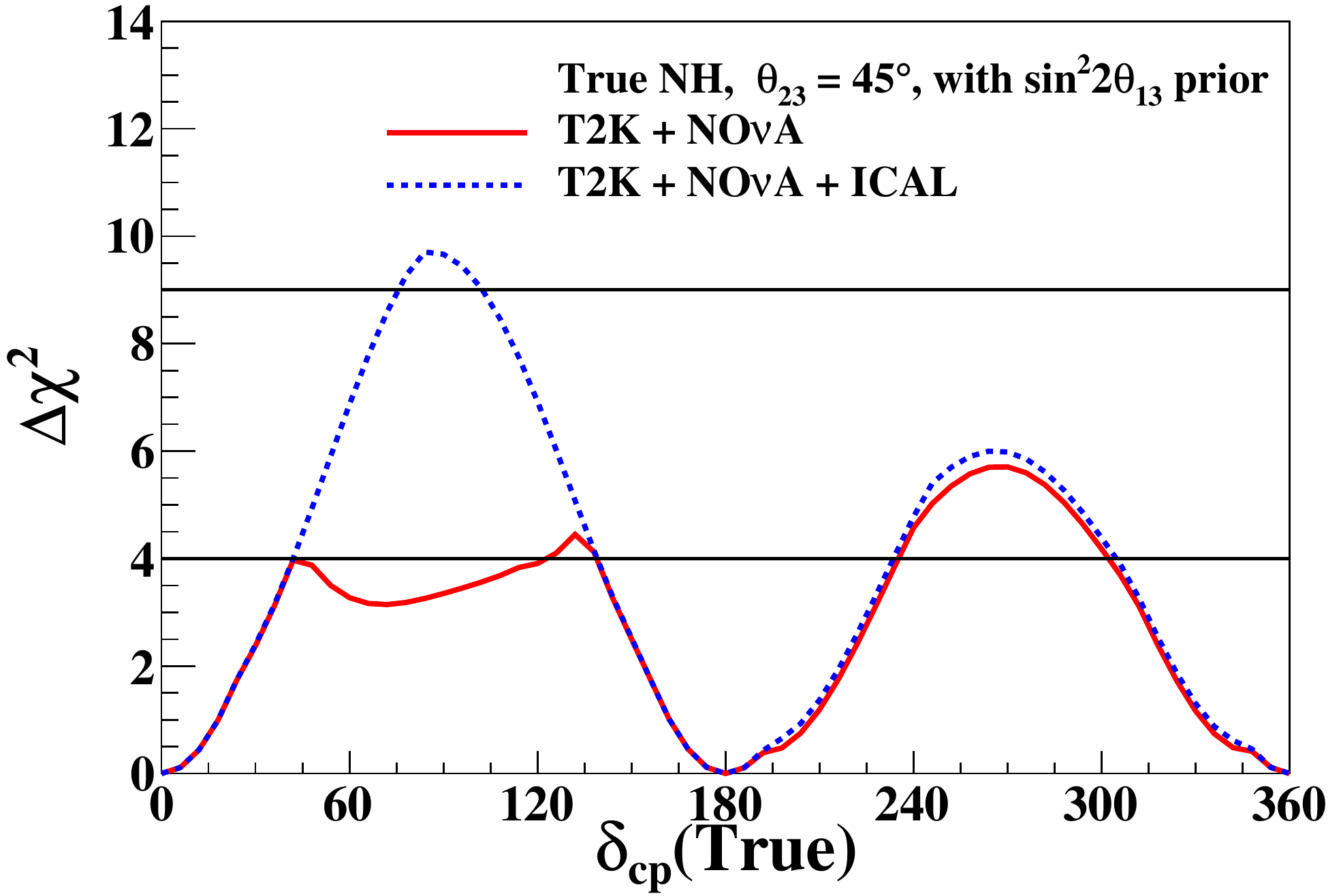}
\caption{Left: Synergies with other experiments that improve the
determination of the mass ordering over the entire $\delta_{CP}$
range. Here the full proposed runs of the long baseline and reactor
experiments are taken, with 10 years' running of ICAL. Right: Synergies
that improve the determination of $\delta_{CP}$.}
\label{fig:synergy}
\end{figure}

The neutrino mass ordering is the centrepiece of ICAL physics. Although
many other experiments such as DUNE and JUNO will also measure this
parameter, ICAL is complementary to these other experiments
because of the separation of the neutrino and anti-neutrino events using
the magnetic field. As a consequence, the sensitivity is independent of
the CP phase $\delta_{CP}$, unlike in other experiments. Since this phase is
currently unknown, the effects of this parameter must be disentangled
from those due to the mass ordering in the other experiments. This can
only be achieved for a fraction of the total range of $\delta_{CP}$. This
feature of ICAL can also be used to obtain synergies with other
experiments. For instance, the NO$\nu$A or ESS$\nu$SB sensitivity
to the mass ordering depends significantly on $\delta_{CP}$; this
sensitivity will be greatly improved on combining with ICAL data;
see Fig.~\ref{fig:synergy}. The solid line shows the sensitivity of
NO$\nu$A to the mass ordering: the ``data" is generated with the normal
ordering and then fitted assuming the inverted ordering, to obtain the
minimum value of $\chi^2$. The results are plotted as a function of the
true value of $\delta_{CP}$, and marginalised over all oscillation
parameters, including $\delta_{CP}$. It is seen that NO$\nu$A has very
little sensitivity to the mass ordering for values of $\delta_{C} <
180^\circ$. Hence the reach of NO$\nu$A alone for the neutrino
mass ordering is very sensitive to the actual value of $\delta_{CP}$.
The sensitivity in this small-$\delta_{CP}$ region improves somewhat
(dotted lines) on including reactor (especially T2K) data; and this
improves dramatically (dashed lines) on including the ICAL information.

Fig.~\ref{fig:synergy} also shows that inclusion of ICAL data will
increase the sensitivity to $\delta_{CP}$ in regions of the
$\delta_{CP}$ plane which earlier had poor sensitivity to this
parameter.

\subsection{Other measurements at ICAL}
\label{sec:other}

\begin{figure}[htp]
\centering
\includegraphics[width=0.48\textwidth]{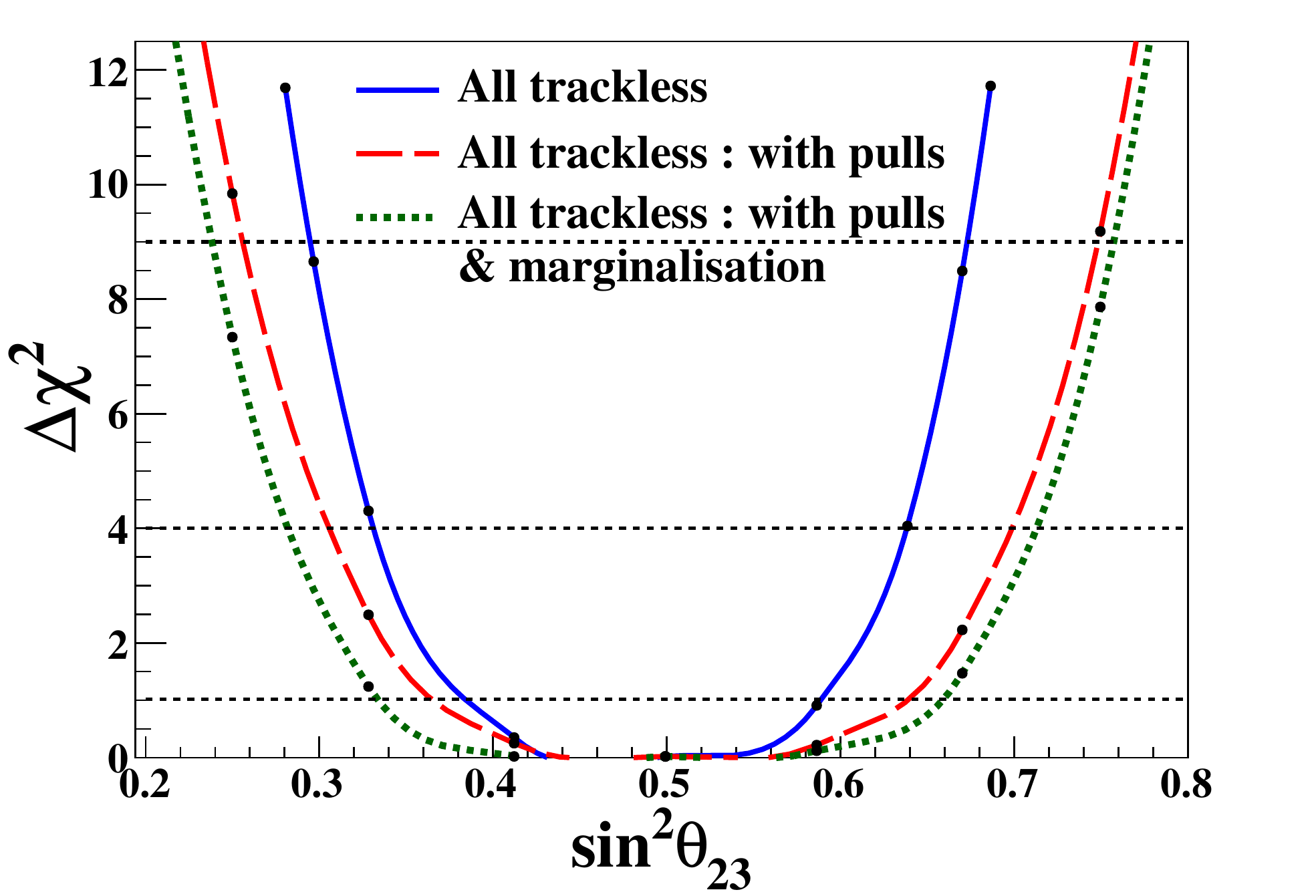}
\caption{Sensitivity to $\sin^2\theta_{23}$ from a simulations analysis
of ``trackless" events. See text for details.}
\label{fig:trackless}
\end{figure}

Electrons from CC interactions of electron neutrinos (and
anti-neutrinos) will also produce showers in ICAL. It is not possible to
separate the trackless events arising from neutral current interactions
of all flavours of neutrinos in ICAL from these electron induced events.
Low energy CC muon events where the muon track is not
reconstructed are an additional contamination. It turns out that the
trackless events also yield information \cite{Chacko:2019wwm} on the
2--3 neutrino mixing parameters, see Fig.~\ref{fig:trackless}. Here, the
solid line indicates the sensitivity to $\sin^2\theta_{23}$ from just
the trackless events at ICAL, when all other oscillation parameters are
kept fixed. The inclusion of systematic uncertainties (dashed line) and
further marginalising over the remaining oscillation parameters (dotted
line) reduces the sensitivity, which remains significant.

\begin{figure}[htp]
\centering
\includegraphics[width=0.4\textwidth,height=0.3\textwidth]{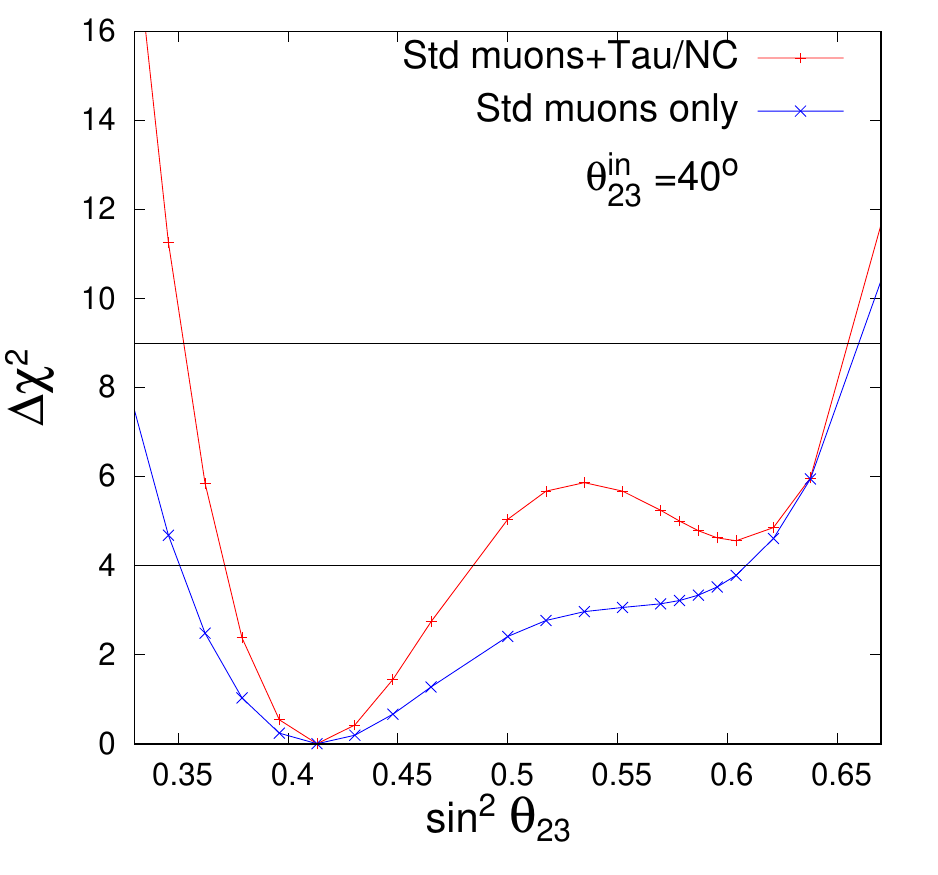}
\hspace{1cm} 
\includegraphics[width=0.4\textwidth,height=0.3\textwidth]{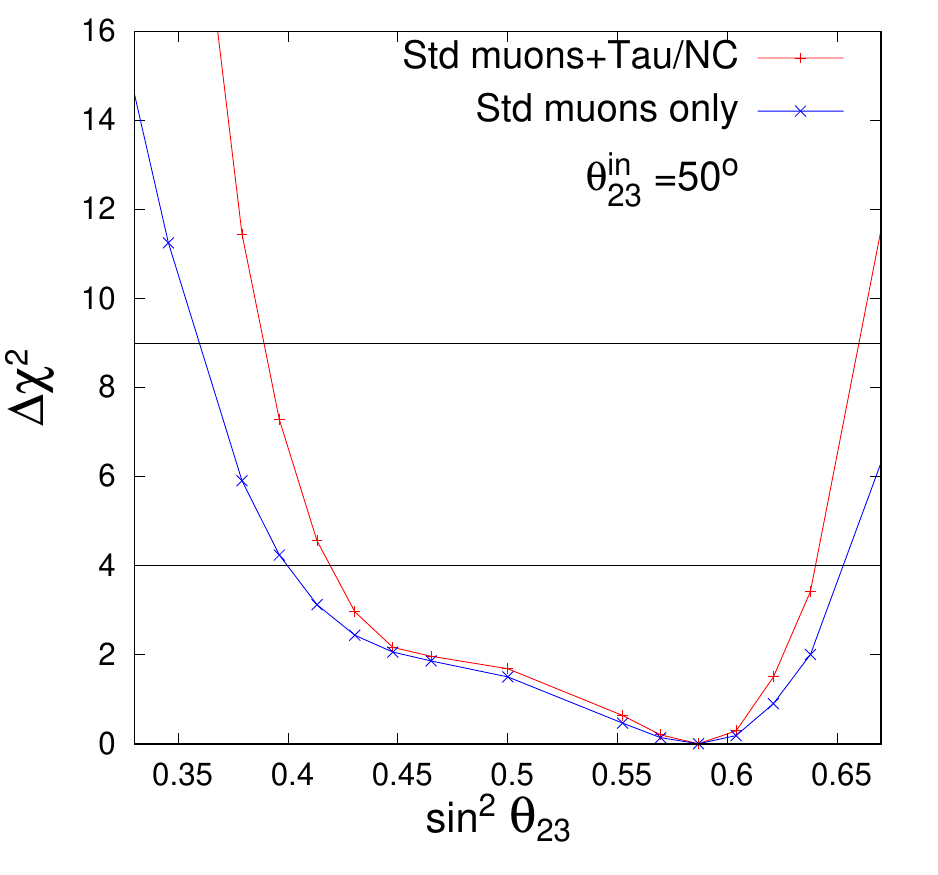}
\caption{Sensitivity to the octant of $\sin^2\theta_{23}$ from a
combined simulations analysis of CC muon and tau events, for an input
value of $\sin^2\theta_{23} = 40^\circ$ (left) and $50^\circ$ (right).}
\label{fig:octanttau}
\end{figure}

Finally, charged tau are produced in CC interactions of tau neutrinos
with the detector. Since atmospheric tau neutrinos are suppressed by a
factor of about $10^5$ compared to atmospheric muon neutrinos,
significant yield of tau neutrino events are a direct signature of
neutrino oscillations, dominantly of the atmospheric muon neutrinos. A
simulations study of the tau neutrino-induced CC events at ICAL, with
the resulting taus decaying hadronically 
\cite{Senthil:2022tmj}, are also sensitive to neutrino
oscillations. Since the threshold for CC tau production is 3.5 GeV,
such events are visible as an excess over the neutral current (NC)
background although they cannot be classified event-by-event. An
analysis of the combined NC and CC$\tau$ events shows sensitivity to
both the 2--3 oscillation parameters $\theta_{23}$ and (the magnitude
of) $\Delta m_{31}^2$. Since CC muon events arise from the {\em same}
atmospheric neutrino fluxes, several systematic uncertainties are the
same, and in fact, simulation studies show that a combined study of muon
and tau CC events improves the sensitivity to the 2--3 oscillation
parameters, in particular, the unknown octant of the
mixing angle $\theta_{23}$, as seen in Fig.~\ref{fig:octanttau}.

There are many other studies of the sensitivity of ICAL to cosmic ray
muons, to BSM physics such as neutrino decays and sterile neutrinos,
non-standard interactions and Lorentz invariance violations, and the
ability to search for dark matter (WIMPS) and exotics such as monopoles.

\section{Prototypes and the mini-ICAL detector}

\begin{figure}[b!]
\centering
\includegraphics[width=0.4\textwidth,height=0.4\textwidth]{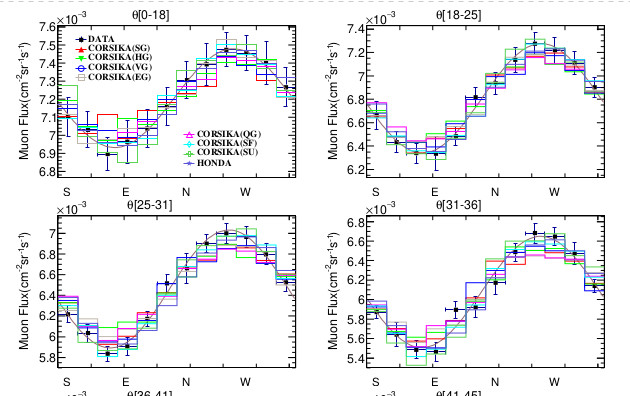}
\hspace{1cm} 
\includegraphics[width=0.4\textwidth,height=0.3\textwidth]{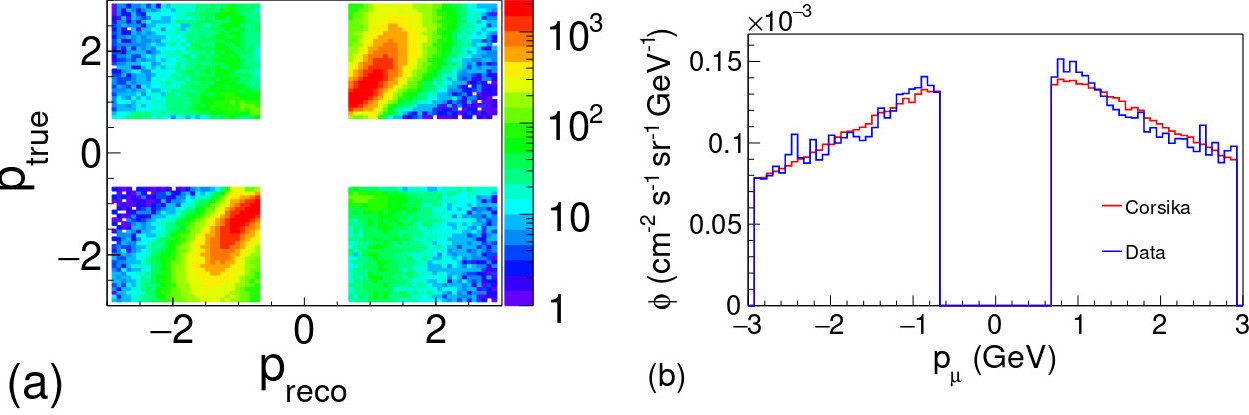}
\caption{Left: Measured east-west asymmetry of cosmic ray muon flux at
the Madurai RPC stack. Right: cosmic muon spectrum from
the mini-ICAL detector at Madurai.}
\label{fig:cosmic}
\end{figure}

Cosmic muons have been observed at several RPC stacks that have been built
for calibration and testing.  One such stack at Madurai in South India
has measured \cite{Pethuraj:2019ryf} the cosmic muon flux as a function
of both the zenith angle and the azimuthal angle and has also measured
the east-west asymmetry of cosmic ray muons; see Fig.~\ref{fig:cosmic}.

\begin{figure}[htp]
\centering
\includegraphics[width=0.6\textwidth]{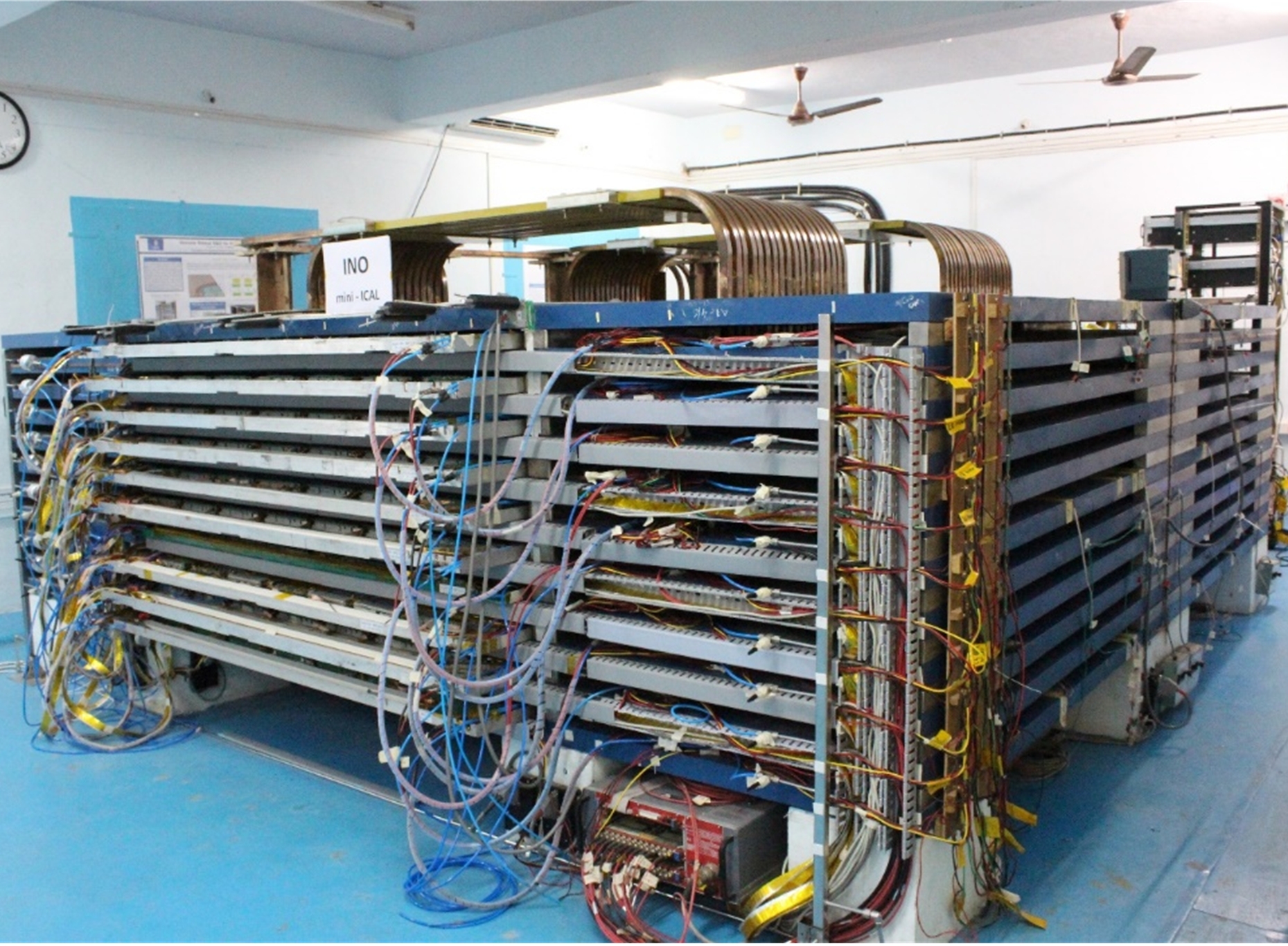}
\caption{The prototype mini-ICAL detector at Madurai, South India.}
\label{fig:miniICAL}
\end{figure}

The mini-ICAL detector; see Fig.~\ref{fig:miniICAL}, is a $4 \times 4$
m$^2$ 85 ton scaled model with 11 iron layers that has been constructed
two years ago. After testing and calibration, the cosmic muon spectrum
has been measured with this detector and is consistent with theoretical
expectations; in addition, the $\mu^-$ and $\mu^+$ events have been separately
observed \cite{Apoorva:2019} and are shown as the spectra with positive and
negative momenta respectively in Fig.~\ref{fig:cosmic}.

\section{Conclusion}
Simulations studies show that ICAL has excellent physics potential,
especially to precisely measure the 2--3 neutrino mixing parameters as
well as the neutrino mass ordering. Detailed studies of charged current
muon events arising from interactions of atmospheric muon neutrinos have
been performed; studies of CC electron neutrino and tau neutrino events
are on-going. Separately, simulations studies have also examined the
sensitivity of ICAL to beyond-the-standard-model (BSM) physics, exotic
particles such as WIMPS and monopoles, etc. ICAL will be complementary
in its reach to other detectors around the world such as DUNE and JUNO.
Finally, ICAL is a completely indigenous detector with excellent R\&D on
RPC design and construction including closed-loop gas system and
automation by industry. In addition, the detector electronics (analog and
digital front end), including chip design and development is also
complete. A 700 ton large scale $8 \times 8 \times 2$ m$^3$ engineering
module is also to be constructed shortly.

In parallel, research is going on into neutrinoless double beta decay
experiments using tin bolometry, and dark matter experiments, which are also
proposed to be housed at INO. While it is expected that INO,
when built, will galvanise students as well as local industry due to
cutting-edge technology transfer, there have been delays in obtaining
some technical clearances for the underground lab construction.  The INO
collaboration is looking forward to going ahead soon with this ambitious
project.

\section*{Acknowledgements}
I thank the conference organisers and the members of the INO
collaboration for this opportunity to present the physics reach and
status report of INO.



%




\bibliography{inotalk.bib}

\begin{thebibliography}{10}
\providecommand{\url}[1]{\texttt{#1}}
\providecommand{\urlprefix}{URL }
\expandafter\ifx\csname urlstyle\endcsname\relax
  \providecommand{\doi}[1]{doi:\discretionary{}{}{}#1}\else
  \providecommand{\doi}{doi:\discretionary{}{}{}\begingroup
  \urlstyle{rm}\Url}\fi
\providecommand{\eprint}[2][]{\url{#2}}

\bibitem{Pontecorvo:1957cp}
B.~Pontecorvo,
\newblock \emph{{Mesonium and anti-mesonium}},
\newblock Sov. Phys. JETP \textbf{6}, 429 (1957).

\bibitem{Maki:1962mu}
Z.~Maki, M.~Nakagawa and S.~Sakata,
\newblock \emph{{Remarks on the unified model of elementary particles}},
\newblock Prog. Theor. Phys. \textbf{28}, 870 (1962),
\newblock \doi{10.1143/PTP.28.870}.

\bibitem{Wolfenstein:1977ue}
L.~Wolfenstein,
\newblock \emph{{Neutrino Oscillations in Matter}},
\newblock Phys. Rev. D \textbf{17}, 2369 (1978),
\newblock \doi{10.1103/PhysRevD.17.2369}.

\bibitem{Mikheev:1986wj}
S.~P. Mikheev and A.~Y. Smirnov,
\newblock \emph{{Resonant amplification of neutrino oscillations in matter and
  solar neutrino spectroscopy}},
\newblock Nuovo Cim. C \textbf{9}, 17 (1986),
\newblock \doi{10.1007/BF02508049}.

\bibitem{Bilenky:2016pep}
S.~Bilenky,
\newblock \emph{{Neutrino oscillations: From a historical perspective to the
  present status}},
\newblock Nucl. Phys. B \textbf{908}, 2 (2016),
\newblock \doi{10.1016/j.nuclphysb.2016.01.025},
\newblock \eprint{1602.00170}.

\bibitem{Esteban:2020cvm}
I.~Esteban, M.~C. Gonzalez-Garcia, M.~Maltoni, T.~Schwetz and A.~Zhou,
\newblock \emph{{The fate of hints: updated global analysis of three-flavor
  neutrino oscillations}},
\newblock JHEP \textbf{09}, 178 (2020),
\newblock \doi{10.1007/JHEP09(2020)178},
\newblock \eprint{2007.14792}.

\bibitem{ICAL:2015stm}
S.~Ahmed \emph{et~al.},
\newblock \emph{{Physics Potential of the ICAL detector at the India-based
  Neutrino Observatory (INO)}},
\newblock Pramana \textbf{88}(5), 79 (2017),
\newblock \doi{10.1007/s12043-017-1373-4},
\newblock \eprint{1505.07380}.

\bibitem{GEANT4:2002zbu}
S.~Agostinelli \emph{et~al.},
\newblock \emph{{GEANT4--a simulation toolkit}},
\newblock Nucl. Instrum. Meth. A \textbf{506}, 250 (2003),
\newblock \doi{10.1016/S0168-9002(03)01368-8}.

\bibitem{Honda:2011nf}
M.~Honda, T.~Kajita, K.~Kasahara and S.~Midorikawa,
\newblock \emph{{Improvement of low energy atmospheric neutrino flux
  calculation using the JAM nuclear interaction model}},
\newblock Phys. Rev. D \textbf{83}, 123001 (2011),
\newblock \doi{10.1103/PhysRevD.83.123001},
\newblock \eprint{1102.2688}.

\bibitem{Casper:2002sd}
D.~Casper,
\newblock \emph{{The Nuance neutrino physics simulation, and the future}},
\newblock Nucl. Phys. B Proc. Suppl. \textbf{112}, 161 (2002),
\newblock \doi{10.1016/S0920-5632(02)01756-5},
\newblock \eprint{hep-ph/0208030}.

\bibitem{Chacko:2019wwm}
A.~Chacko, D.~Indumathi, J.~F. Libby and P.~K. Behera,
\newblock \emph{{First simulation study of trackless events in the INO-ICAL
  detector to probe the sensitivity to atmospheric neutrinos oscillation
  parameters}},
\newblock Phys. Rev. D \textbf{102}(3), 032005 (2020),
\newblock \doi{10.1103/PhysRevD.102.032005},
\newblock \eprint{1912.07898}.

\bibitem{Senthil:2022tmj}
R.~T. Senthil, D.~Indumathi and P.~Shukla,
\newblock \emph{{A simulation study of tau neutrino events at the ICAL detector
  in INO}}  (2022),
\newblock \eprint{2203.09863}.

\bibitem{Pethuraj:2019ryf}
S.~Pethuraj, G.~Majumder, V.~M. Datar, N.~K. Mondal, K.~C. Ravindran and
  B.~Satyanarayana,
\newblock \emph{{Measurement of azimuthal dependent muon flux by 2 m x 2 m RPC
  stack at IICHEP-Madurai}},
\newblock Exper. Astron. \textbf{49}(3), 141 (2020),
\newblock \doi{10.1007/s10686-020-09655-y},
\newblock \eprint{1905.00739}.

\bibitem{Apoorva:2019}
A.~Bhatt,
\newblock \emph{{ Measurement of Atmospheric Muons at IICHEP in Madurai, for
  better estimation of Neutrino Fluxes at INO Site in Theni}}  (2019).

\end{thebibliography}

\nolinenumbers

\end{document}